\documentclass[preprint,proceedings]{rmaa}

\newcommand{\D}{\discretionary{}{}{}}

\SetYear{2009}
\SetConfTitle{The Interferometric View on Hot Stars}

\title{Long baseline interferometry: a promising tool for multiplicity investigations of massive stars}

\author{
  M. De Becker\altaffilmark{1}}

\altaffiltext{1}{Institut d'Astrophysique et G\'eophysique, University of Li\`ege, 17 all\'ee du 6 Ao\^ut, B\^at. B5c, 4000 Li\`ege (Sart-Tilman), Belgium
  (debecker{}@astro.\D{}ulg.\D{}ac.\D{}be)}

\suppressfulladdresses

\listofauthors{M. De Becker}

\indexauthor{De Becker, M.}

\addkeyword{Stars: Binaries}
\addkeyword{Stars: Early-type}

\begin{document}

\maketitle 

\section{Multiplicity studies}

The issue of the multiplicity of massive stars can be addressed through the investigation of radial velocity variations in spectral lines, or by the detection of blended asymmetric line profiles. However, the absence of such signatures should by no mean be interpreted as an evidence that the star is single. The non-detection of the spectral line of the secondary can indeed be due to a high luminosity ratio of the two components of the binary. The lack of significant radial velocity variation can also be due to a very low inclination angle of the orbit, a poor sampling of an eccentric orbit, or even a very long orbital period.

On the other hand, interferometry can also be used to investigate stellar multiplicity, but most of the Galactic massive stars studied so far are located at distances of a few kpc. Considering the angular resolution of present facilities such as the VLTI ($\sim$\,mas), typical orbital separations of a few AUs are accessible. Applying Keplers laws, this translates into periods of at least a few years or so. A large inclination angle and a large eccentricity may compromise further the resolution of binaries, at least during a significant fraction of their orbit. Interferometry is therefore mostly efficient at investigating long period binaries.

Both approaches present therefore different limitations, and should be considered as complementary. Long baseline interferometry is therefore expected to open a new part of the parameter space of massive binaries (see also Sana \& Le Bouquin in this volume).

\section{Physical processes}

Beside the determination of orbital parameters, one of the main interest of such studies is to provide crucial information useful to investigate physical processes related to the binarity of massive stars.

\paragraph{Thermal X-ray emission}  The interaction between stellar winds in a massive binary is likely to generate strong hydrodynamic shocks, resulting in a substantial heating of the post-shock plasma ($\sim$\,10$^7$\,K). Such a hot plasma will produce a thermal X-ray spectrum determined by the physical properties of the colliding-wind region, notably dependent on the binary separation. The modelling of the hydrodynamics of colliding winds, and of the related X-ray emission, requires therefore an accurate determination of the orbital parameters through an adequate multiplicity study (Sana et al.\,2004, De Becker et al.\,2006). Such information is also needed in order to define strategies for observations with e.g. XMM-Newton aiming at studying the X-ray emission from colliding-wind massive binaries.
\paragraph{Dust production} Some Wolf-Rayet binaries display the signature of a significant dust production process in the wind-wind interaction region (see e.g. De Becker et al. in this volume, and references therein). The modelling of the dust formation process requires an accurate determination of orbital parameters as well.
\paragraph{Particle acceleration} Several colliding-wind massive binaries turn out to be particle accelerators: evidence for the presence of relativistic electrons has been found in the case of more than 30 massive systems (O-type and Wolf-Rayet, see De Becker\,2007). These relativistic electrons point to the existence of an efficient acceleration process at work in massive star environments. The current standard model assumes that this acceleration takes mainly place close to the wind-wind interaction region (Pittard \& Dougherty\,2006, De Becker\,2007).\\

Multiplicity constitutes a key point in the investigation of the physics of massive stars. Observation campaigns with the VLTI have been initiated in order to investigate the multiplicity of a few massive stars (HD\,167971, WR\,106, WR\,112). The orbital parameters are intended to be used in the context of the investigation of the physics of colliding winds.

\end{document}